\documentstyle[aps,prl,amssymb,preprint]{revtex}
\begin{document}
\draft

\title{A unified approach to Hamiltonian systems, Poisson systems, gradient
systems, and systems with Lyapunov functions and/or first integrals}

\author{Robert I. McLachlan}
\address{Mathematics Department, Massey University, 
Palmerston North, New Zealand}
\author{G.R.W. Quispel}
\address{Department of Mathematics, LaTrobe University, 
Bundoora, Melbourne 3083, Australia}
\author{Nicolas Robidoux}
\address{Mathematics Department, Massey University, 
Palmerston North, New Zealand}


\maketitle

\def\smallmatrix{\null\,\vcenter\bgroup \baselineskip=7pt
 \ialign\bgroup\hfil$\scriptstyle{##}$\hfil&&$\;$\hfil
 $\scriptstyle{##}$\hfil\crcr}
\def\endsmallmatrix{\crcr\egroup\egroup\,}
\def\smatrix#1{\smallmatrix #1 \endsmallmatrix}
\def\spmatrix#1{\left( \smallmatrix #1 \endsmallmatrix \right)}

\begin{abstract}
Systems with a first integral (i.e., constant of motion)
or a Lyapunov function can be written as ``linear-gradient systems''
$\dot x= L(x)\nabla V(x)$ for an appropriate matrix function $L$,
with a generalization to several integrals or Lyapunov functions.
The discrete-time analogue, $\Delta x/\Delta t = L \overline\nabla V$ where
$\overline\nabla$ is a ``discrete gradient,'' preserves $V$ as an integral
or Lyapunov function, respectively.
\end{abstract}

\narrowtext

\section{Introduction}

Integrals and Lyapunov functions---quantities that are conserved
or dissipated, respectively---are fundamental in dynamics.
They severely constrain the system's evolution and can be used
to establish stability. There is no universal method to find
such quantities, but if they are known (e.g., on physical grounds)
we show that the system can be presented in a universal form which
makes the conservation/dissipation property manifest. Although elementary,
this result is very general and will find many applications: here we
use it to preserve the conservation/dissipation property under 
time discretization.

We start with the definition and an example of each of 
the classes of systems covered in this Letter.
\begin{itemize}
\item[(i)] Hamiltonian systems.

Hamiltonian systems are ubiquitous in physics \cite{arnold}.
They have the form
$\dot x = J \nabla V(x)$, $x \in {\mathbb R}^{2n}$,
where $V(x)$ denotes the Hamiltonian function, and
$ J:= \spmatrix{0 & {\rm Id}\cr -{\rm Id} & 0}$,
where Id denotes the identity matrix in ${\mathbb R}^n$.

\noindent{\em Example 1.} A simple Hamiltonian system is the
pendulum \cite{arnold}
$\dot x_1 = x_2$, $\dot x_2 = - \sin (x_1)$;
here $n=1$ and $V(x_1, x_2) = {1\over 2} x^2_2 - \cos (x_1)$.

\item[(ii)] Poisson systems.

Poisson systems also occur very frequently in phys-ics (\cite{arnold},
App. 14).
They have the form
$\dot x = \Omega (x) \nabla V(x)$, $x \in {\mathbb R}^n$,
where $V(x)$ again denotes the Hamiltonian function and the Poisson structure
$\Omega (x)$ is an antisymmetric matrix
($\Omega^t (x) = - \Omega (x)$),
satisfying the Jacobi identity
$\Omega_{jk} \partial_k \Omega_{\ell m} + \Omega_{\ell k} \partial_k
\Omega_{mj} + \Omega_{mk} \partial_k \Omega_{j \ell} = 0.$

\noindent{\em Example 2.}
The equations of motion of a free rigid body with moments of
inertia $I_1$, $I_2$, and $I_3$ form 
a Poisson system with angular momentum $x\in{\mathbb R}^3$ and
Poisson structure
\begin{equation}
\Omega(x) = \left(\begin{array}{rrr}
0 & x_3 & -x_2 \\
-x_3 & 0 & x_1 \\
x_2 & -x_1 & 0 \\
\end{array}\right)
\end{equation}
and Hamiltonian 
$ V(x) = {1\over2}\sum_{i=1}^3 x_i^2/I_i$.
(Actually this is an example of a so-called Lie-Poisson structure, in
which $\Omega(x)$ is a linear function.)

\item[(iii)] Systems with a first integral.

The ordinary differential equation (ODE)
$\dot x = f(x)$, $x\in{\mathbb R}^n$,
is said to have the first integral $V$ if
$dV(x)/dt = 0$.

\noindent{\em Example 3.}  A Lotka-Volterra system \cite{gr-etal}.
The ODE
\begin{equation}
\label{ex3}
\dot x_1 = e^{x_3},\ 
\dot x_2 = e^{x_1} + e^{x_3},\ 
\dot x_3 = Be^{x_1} + e^{x_2},
\end{equation}
where $B$ is a parameter, possesses the integral
\begin{equation}
\label{Vex3}
V(x_1, x_2, x_3) = e^{x_2 - x_1} + B (x_2 - x_1) - x_3.
\end{equation}

\item[(iv)] Gradient systems.

Gradient systems arise, e.g.,  in dynamical systems theory
\cite{hi-sm}.  They are described by
$\dot x = - \nabla V (x)$, $x \in {\mathbb R}^n$.

\noindent{\em Example 4.} The system 
$\dot x_1 = -2x_1 (x_1 - 1)(2x_1 - 1)$, 
$\dot x_2 = -2 x_2$ is a gradient system \cite{hi-sm}, with 
$n = 2$ and $V(x_1, x_2) = x^2_1 (x_1 - 1)^2 + x^2_2$.

\item[(v)] Systems with a Lyapunov function.

The ODE
$\dot x = f(x)$, $x \in {\mathbb R}^n$,
is said to possess the Lyapunov function $V$ if
$dV(x)/dt \leq 0$.
These functions were introduced by Lyapunov \cite{lyapunov}
and are a crucial ingredient
of his direct or second method in the study of dynamical stability
\cite{gu-ho,ro-ha-la}. 

\noindent{\em Example 5.} \cite{sanchez}
\begin{equation}
\label{ex5}
\dot x_1 = -x_2 - x_1^3, \quad
\dot x_2 = x_1 - x_2^3
\end{equation}
has the Lyapunov function
\begin{equation}
\label{Vex5}
V (x_1, x_2) = x_1^2 + x_2^2.
\end{equation}
\end{itemize}

What do the above five classes of dynamical systems have in common?
A preliminary answer would be that they all possess a function
$V(x)$ such that $dV(x)/dt \leq 0$.  That is, classes
(i), (ii) and (iii) each possess a function $V(x)$ such that
$dV(x)/dt \equiv 0$,
and classes (iv) and (v) each possess a function
$V(x)$ s.t. $dV(x)/dt \leq 0$.

In Section 2 we announce the result that classes of systems (i) to (v)
have even more in common: under some mild technical assumptions, they can
all be written as special cases of the novel class of ``linear-gradient
systems.''
In Section 3 we show how these linear-gradient systems can be integrated
numerically in such a way that $V(x)$ is constant or non-increasing,
as appropriate.

An extended version of this work, including proofs of the results presented
here, is given in \cite{mc-qu-ro}.

\section{``Linear-gradient systems''}

Our main result is the following:

\noindent{\bf  Theorem 1.} { \em Let the ODE
\begin{equation}
\label{ode}
\dot x = f(x), \qquad f \in C^r,
\end{equation}
possess a $C^{r+1}$ Morse function $V(x)$, where
\begin{itemize}
\item[(a)]  \qquad
$\displaystyle \frac{dV}{dt} = 0$, i.e., $V$ is an integral; or
\item[(b)]  \qquad
$\displaystyle \frac{dV}{dt} \leq 0$, i.e., $V$ is a
(weak) Lyapunov function; or
\item[(c)]  \qquad
$\displaystyle \frac{dV}{dt} < 0 $
where $f(x)\ne 0$, i.e., $V$ is a strong Lyapunov function.
\end{itemize}
Then for all $\{x \mid
\nabla V (x) \not= 0\}$ there exists a locally bounded $C^r$ matrix $L(x)$
such that the ODE (\ref{ode}) can be rewritten in the linear-gradient form
\begin{equation}
\label{linear-gradient}
\dot x = L(x) \nabla V (x),
\end{equation}
where
\begin{itemize}
\item[(a)]  \qquad $L(x)$ is an antisymmetric matrix; resp.
\item[(b)]  \qquad $L(x)$  is a negative semidefinite matrix; resp.
\item[(c)]  \qquad $L(x)$ is a negative definite matrix.
\end{itemize}
}


\noindent{\bf Remarks:}

\begin{itemize}

\item[1.] A Morse function is a function whose
critical points are all nondegenerate. A negative semidefinite matrix $L$ is
a matrix such that $v^t L v \leq 0$ for all vectors $v$. A negative
definite matrix $L$ is a matrix such that $v^t L v < 0$ for all non-zero
vectors $v$.

\item[2.] Under a coordinate transformation $x \mapsto C(x)$ we have
$L(x) \mapsto \tilde L (x)
:= d C(x) L(x) (dC(x))^t$.
This implies that the theorem is invariant under coordinate transformations,
because $\tilde L$ is antisymmetric, negative
semidefinite, resp. negative definite iff $L$ is.

\item[3.] The theorem has a converse: if an ODE is in linear-gradient
form (\ref{linear-gradient}) with $L$ antisymmetric, 
resp. negative semidefinite, resp.
negative definite, then $V$ is an integral, resp. weak Lyapunov function,
resp. strong Lyapunov function.

\item[4.] If the sign of $dV/dt$ (zero, nonpositive, or negative)
depends on $x$, then $L$ can be chosen to be antisymmetric, negative
semidefinite, or negative definite respectively, depending on $x$.
The type of representation is not unique: at points where $dV/dt=0$,
$L$ can be chosen to be either antisymmetric or negative semidefinite.

\item[5.] A particular $L(x)$ satisfying the requirements of the theorem
is
\begin{equation}
L_{ij}(x) = {f_i v_j - v_i f_j +  \delta_{ij}\sum f_k v_k\over \sum v_k^2}
\end{equation}
where $v_j = \partial V/\partial x_j$.
However, $L$ in (\ref{linear-gradient}) yielding (\ref{ode}) is not unique.
In particular, under further mild technical
conditions, there is an $L$ which extends smoothly through critical
points of $V$.

\item[6.] The fact that all systems with an integral can be written
in the skew-gradient form $\dot x = L(x)\nabla V(x)$ was, as
far as we know, first published in \cite{qu-ca}.
The general case is new, although the special
case of the converse with  $L(x)$ symmetric negative definite is well known
and forms the subject of ``generalized gradient systems'' in
dynamical systems \cite{hi-sm}.
\end{itemize}

\noindent The constructive proof of Theorem 1 is given in \cite{mc-qu-ro}.
We now give some illustrative examples of the above theorem.

\noindent{\em Example 6.} Particle in 1D with friction \cite{hi-sm}.
Consider the ODE
\begin{equation}
\label{ex6}
\dot x_1 =  x_2, \quad 
\dot x_2 =  -\frac{\partial f(x_1)}{\partial x_1} - \alpha x_2,
\end{equation}
where $\alpha \geq 0$ is a coefficient of friction, and $f$ is a potential
function.  Eq. (\ref{ex6}) has the energy 
$ V (x_1, x_2) = \frac{1}{2} x^2_2 + f(x_1)$
as a Lyapunov function,
and can be written in the linear-gradient form (\ref{linear-gradient}) as 
\begin{equation}
\left(\begin{array}{c} \dot x_1\\ \dot x_2 \end{array}\right) =
\left(\begin{array}{cc} 0 & 1 \\ -1 & -\alpha \end{array}\right) \nabla V
(x_1, x_2).
\end{equation}
For $\alpha = 0$, the system is conservative, and the matrix $L$ is
antisymmetric (case (a) above).  For $\alpha > 0$, the system is
dissipative, $V$ is a (weak) Lyapunov function, and
$L(x)$ is negative semidefinite (case (b) above;  cf \S9.4 of
\cite{hal-ko}).

\noindent{\em Example 7.}  An averaged system in wind-induced
oscillation \cite{gu-ho}.
Consider the system
\begin{eqnarray}
\label{ex7}
\dot x_1 &=& -\zeta  x_1 - \lambda x_2 + x_1 x_2 \nonumber \\
\dot x_2 &=& \lambda x_1 - \zeta  x_2 + \frac{1}{2} (x_1^2 - x_2^2).
\end{eqnarray}
Here $\zeta  \geq 0$ is a damping factor and $\lambda$ is a detuning
parameter.  Guckenheimer and Holmes \cite{gu-ho}
remark that for $\zeta  = 0$, Eq. (\ref{ex7}) has the
Hamiltonian
\begin{equation}
V(x_1, x_2) = -\frac{\lambda}{2} \left(x_1^2 + x_2^2\right) +
\frac{1}{2} \left(x_1 x_2^2 - \frac{x^3_1}{3}\right),
\end{equation}
and for $\lambda = 0$, Eq. (\ref{ex7}) is a gradient system with
\begin{equation}
V (x_1, x_2) = \frac{\zeta }{2} \left(x_1^2 + x_2^2\right) + \frac{1}{2}
\left(\frac{x^3_2}{3} - x_1^2 x_2\right).
\end{equation}
We now show that, for all allowed values of $\zeta $ and $\lambda$, Eq.
(\ref{ex7}) can
be written in linear-gradient form.  To this end denote $\zeta  = \rho \cos
(\theta)$
and $\lambda = \rho \sin (\theta)$.  Then Eq. (\ref{ex7}) can be written in
linear-gradient form $\dot x = L\nabla V$
with
\begin{equation}
L = \left(\begin{array}{cc} -\cos (\theta) &-\sin (\theta) \\
\sin(\theta) &-\cos(\theta) \end{array}\right),
\end{equation}
and
$$
V(x_1, x_2) = \frac{1}{2} \rho \left(x^2_1 + x^2_2\right) -
\frac{1}{2} \sin(\theta) \left(x_1 x_2^2 - \frac{x_1^3}{3}\right) 
$$
\begin{equation}
\label{Vex7}
\qquad{}+\frac{1}{2} \cos (\theta) \left(\frac{x^3_2}{3} - x_1^2 x_2\right).
\end{equation}
Note that for the matrix $L$ in this example we have
$ v^t Lv = - \cos (\theta) \vert v^2 \vert$.
Therefore, in the physical regime (where $\zeta  \geq 0$ and hence
$\cos (\theta)
\geq 0$) either the matrix $L$ is antisymmetric and $V$ is an integral (for
$\cos (\theta) = 0$), or $L$ is negative definite and $V$ is a strong
Lyapunov function (for $\cos (\theta) > 0$).
(Note that for $\lambda = \zeta  = 0$, we have $\rho = 0$ and we are free to
choose $\theta$.  In this limit, the system possesses an integral
\begin{equation}
V_1 (x_1, x_2) = \frac{1}{2} \Big(x_1 x_2^2 - \frac{x_1^3}{3}\Big),
\end{equation}
as well as a Lyapunov function
\begin{equation}
V_2 (x_1, x_2) = \frac{1}{2} \Big(\frac{x_2^3}{3} - x_1^2 x_2\Big),
\end{equation}
and $V$ given by Eq. (\ref{Vex7}) represents an arbitrary linear 
combination of these two functions).

\noindent{\em Example 8.}
Reproduced from \cite{qu-ca}, here
is the linear-gradient form for the ODE (\ref{ex3}) in
Example 3:
\begin{equation}
\left(\begin{array}{c} \dot x_1\\ \dot x_2\\ \dot x_3\end{array}\right) =
\left(\begin{array}{ccc} 0 & 0 &-e^{x_3} \\ 0 & 0 &-e^{x_1}-e^{x_3} \\
e^{x_3} &e^{x_1} + e^{x_3} &0 \end{array}\right) \nabla V,
\end{equation}
where $V$ is given by Eq. (\ref{Vex3}).

\noindent{\em Example 9.}
Here is the linear-gradient form for the ODE (\ref{ex5}) in Example 5:
\begin{equation}
\label{ex9}
\left(\begin{array}{c} \dot x_1\\ \dot x_2 \end{array}\right)  
= 
\left(\begin{array}{cc} 
a & b \\ -b & a \end{array} \right)
\nabla V (x_1, x_2),
\end{equation}
where $a=-(x_1^4+x_2^4)/(x_1^2+x_2^2)$,
$b=-(x_1^2+x_2^2+x_2 x_1^3 - x_1 x_2^3)/(x_1^2+x_2^2),$
and $V(x_1, x_2)$ is given by Eq. (\ref{Vex5}).
Note that the matrix $L$ in (\ref{ex9}) is negative definite.

\section{Discrete gradients and the numerical integration of
linear-gradient systems}

For differential equations whose time evolution has particular structural
properties, such as preservation of symplectic structure, phase
space volume, symmetries, or conserved quantities, it is desirable
to mimic these properties in any numerical integration \cite{st-hu}.
This is particularly useful in long-time integrations.
One can also view the discrete-time analogues as interesting physical
systems in their own right \cite{lee}.

A major application of the linear-gradient formulation 
(\ref{linear-gradient}) is that
it has a simple and elegant discrete-time analogue; moreover,
this analogue is also a universal representation for systems of each
class.

\noindent{\bf Definition 1.} \cite{gonzalez}
{\em Let $V(x)$ be a differentiable
function.  Then $\overline\nabla V(x, x')$ is a {\bf discrete gradient} of
$V$ if it is continuous and}
\begin{eqnarray}
\label{dg}
&&\overline\nabla V (x, x') \cdot (x' - x) = V(x') - V (x), \nonumber \\
&& \overline\nabla V (x, x) = \nabla V (x).
\end{eqnarray}

Discrete gradients are not unique.  Several examples of discrete gradients
are given in \cite{gonzalez,ha-la-le,mc-qu-ro}.

\noindent{\bf Definition 2.} {\it The function $V$ is an integral of the map
$x \mapsto x'$ if $V(x')$ = $V(x)$, $\forall x$.  It is a weak
Lyapunov function if $V(x') \leq V(x), \ \forall x$.  It is a strong
Lyapunov function if $V(x') < V(x)$ for all $x$ such that
$x \ne x'$.}

\noindent{\bf Theorem 2.} \cite{mc-qu-ro}
{\it Let the map $x \mapsto x'$ be defined implicitly by
\begin{equation}
\label{map}
\left(\frac{\Delta x}{\Delta t}=\right)
\frac{x' - x}{\tau} = \tilde L (x, x', \tau)
\overline \nabla V (x, x')
\end{equation}
where $\overline \nabla V$ is any discrete gradient, $\tilde
L$ is a matrix function, and $\tau$ represents a timestep.
Then $V(x)$ is an
integral, resp. weak Lyapunov function, resp. strong  Lyapunov function of
the map if $\tilde L$ is antisymmetric, resp. negative semidefinite,
resp. negative definite.  Conversely, for any map with
such a $V$, and any discrete gradient $\overline\nabla$, at points
such that $\overline\nabla V(x,x')\ne 0$ there exists an $\tilde L$
such that that map takes the form (\ref{map}).
}

It follows that (\ref{map}) is a discrete approximation to the 
linear-gradient system (\ref{linear-gradient}) that preserves integrals, 
resp. Lyapunov functions, provided the method
is consistent, i.e., $\tilde L (x, x, 0) = L (x)$.

Equations similar to (\ref{dg}), (\ref{map}) have appeared in many
energy-conserving schemes for Hamiltonian systems
\cite{ch-hu-ma-mc,gotusso,it-ab,kriksin}, although the first axiomatic
presentation was \cite{gonzalez}, and the first application to
all systems with an integral was \cite{qu-tu}.

\section{Concluding remarks}

\begin{itemize}

\item[(i)] In this Letter, for simplicity we have restricted our discussion
to the case of {\em one} first integral or Lyapunov function.  In
\cite{mc-qu-ro} we show
that an $n$-dimensional ODE with $m\leq n-1$ integrals and/or Lyapunov 
functions
$V_1, \dots, V_m$, can be written in the ``multilinear-gradient'' form
\begin{equation}
\label{multi}
\dot x = L (x) \nabla V_1 \dots \nabla V_m,\quad  x \in {\mathbb R}^n,
\end{equation}
where $L(x)$ is an $(m+1)$-tensor.
Structure-preserving integrators for Eq. (\ref{multi}) have also 
been constructed, generalizing Eq. (\ref{map}).

\item[(ii)] Associated with (multi)linear-gradient systems 
of the form (\ref{multi})
there is also a formulation in terms of a bracket:
\begin{equation}
\frac{df(x)}{dt} = \{f, V_1, \dots, V_m\}_L,
\end{equation}
where the bracket is defined by
\begin{equation}
\label{multib}
\{f_1, \dots, f_{p}\}_L := \sum_{i_1, \dots, i_{p}} L_{i_1, \dots,
i_{p}} \frac{\partial f_1}{\partial x_{i_1}} \dots \frac{\partial
f_{p}}{\partial x_{i_p}}
\end{equation}
where $p=m+1$.
This bracket satisfies the Leibnitz rule in each of its variables:
\begin{eqnarray}
&&\{f_1, \dots, f_{j-1}, \phi (g_1, \dots, g_k), f_{j + 1}, \dots, f_{p
}\}_L \nonumber \\
= &&\sum^k_{i=1} \left(\frac{\partial \phi}{\partial g_i}\right) \{f_1,
\dots, f_{j-1}, g_i, f_{j + 1}, \dots, f_{p}\}_L,
\end{eqnarray}
$j=1,\dots,p$.
Conversely, the tensor $L$ is defined by the fundamental brackets
$ L_{i_1, \dots, i_{p}} = \{x_{i_1}, \dots,$ $ x_{i_{p}}\}_L$.
It follows that $V(x)$ is an integral, resp. weak Lyapunov function, resp.
strong Lyapunov function of the (multi)linear-gradient system (\ref{multi}) iff
$W=0$ $\forall x$, resp. $W\le 0$ $\forall x$, resp. $W<0$ for all $x$
such that $ \vert \nabla V (x) \vert \not= 0$, where $W :=
\{V,V_1,\dots,V_m\}_L$.
It  also follows that $V$ is an integral (resp. Lyapunov function) of the
system (\ref{multi}) iff $V_j$ is an integral (resp. Lyapunov function) 
of the system
\begin{equation}
\dot x = \tilde L (x) \nabla V_1 \dots \nabla V_{j-1} \nabla V \nabla
V_{j + 1} \dots \nabla V_m,
\end{equation}
where
\begin{equation}
\tilde L_{i_{1, \dots, i_{j-1}, i_j, i_{j+1}, \dots i_m}} = L_{i_j, \dots,
i_{j-1}, i_1, i_{j + 1}, \dots, i_m}.
\end{equation}
Special cases of the bracket (\ref{multib}) are the Poisson bracket and 
the Nambu bracket \cite{takhtajan}.

\item[(iii)] We hope to address the numerical order of accuracy of the
integrator (\ref{map}) in a forthcoming publication.

\end{itemize}

\noindent{\bf Acknowledgements}

We are indebted to the Marsden Fund of the Royal Society of New Zealand, and
Lottery Science New Zealand, for their financial support.  G.R.W.Q. is also
grateful for a La Trobe University Large Central Grant and an Australian
Research Council Small Grant supporting this project.


\begin{thebibliography}{99}

\bibitem{arnold} V.I. Arnold, {\em Mathematical Methods of Classical
Mechanics}, 2nd ed., Springer-Verlag, New York, 1989.

\bibitem{ch-hu-ma-mc} A.J. Chorin, T.J.R. Hughes, J.E. Marsden, and
M. McCracken, {\em
Comm. Pure Appl. Math. \bf 31} (1978), 205--256.

\bibitem{gonzalez} Oscar Gonzalez, 
  {\em J. Nonlinear Sci. \bf 6}(5), 449--467
  (1996).

\bibitem{gotusso} L. Gotusso, 
{\em Appl. Math. Comput. \bf 17} (1985), 129--136.

\bibitem{gr-etal} B. Grammaticos, J. Moulin-Ollagnier, A. Ramani,
J.-M. Strelcyn, and S. Wojciechowski, {\em Physica A \bf 163} (1990) 683.

\bibitem{gu-ho} J. Guckenheimer and P. Holmes, {\em Nonlinear Oscillations,
Dynamical Systems, and Bifurcations of Vector Fields}, Springer-%
Verlag, New York, 1983.


\bibitem{hal-ko} J.K. Hale and H. Ko\c cak, {\em Dynamics and
Bifurcations}, Springer, Berlin,
1991.
\bibitem{ha-la-le} Amiram Harten, Peter D. Lax, and Bram van Leer,
{\em SIAM Review \bf 25}(1) (1983), 35--61.

\bibitem{hi-sm} Morris Hirsch and Stephen Smale,
{\em Differential equations, dynamical systems, and linear algebra},
Academic Press, New York, 1974.

\bibitem{it-ab} T. Itoh and K. Abe, 
{\em J. Comput. Phys. \bf 77} (1988) 85.

\bibitem{kriksin} Yu.A. Kriksin, 
{\em Zh. Vychisl. Mat. Fiz. \bf 33} (1993), 206--218.

\bibitem{lee} T.D. Lee, 
{\em J. Stat. Phys. \bf 46} (1987)  843--860.

\bibitem{lyapunov} A.M. Lyapunov, {\em The general problem of the stability
of motion}, Taylor and Francis, London, 1992.

\bibitem{mc-qu-ro} R.I. McLachlan, G.R.W. Quispel, and N. Robidoux,
{\em Phil. Trans.  Roy. Soc. A}, submitted.

\bibitem{olver} Olver, P. J., {\em Applications of Lie groups to
Differential Equations}, 2nd ed., Springer-Verlag, New York, 1993.

\bibitem{qu-ca} G.R.W. Quispel and H.W. Capel, 
  {\em Phys. Lett. A \bf 218} (1996), 223-228.

\bibitem{qu-tu} G.R.W. Quispel and G.S. Turner, 
  {\em J. Phys. A \bf 29} (1996), L341--L349.

\bibitem{ro-ha-la} N. Rouche, P. Habets, and M. LaLoy, {\em Stability
theory by Liapunov's direct method}, Springer Verlag, New York, 1977.

\bibitem{sanchez}  D.A. Sanchez, {\em Ordinary Differential Equations
and Stability Theory}, WH Freeman, San Francisco, 1968.

\bibitem{st-hu} A.M. Stuart and A.R. Humphries, {\em Dynamical systems
and numerical analysis}, CUP, 1996.

\bibitem{takhtajan} L. Takhtajan, 
{\em Comm. Math. Phys. \bf 160} (1994), 295--315.

\end{thebibliography}
\end{document}